\documentclass[12pt,preprint]{aastex}

\slugcomment{Accepted for publication in the Astrophysical Journal}

\shorttitle{Nuclear Gas Clouds in Centaurus A}
\shortauthors{Bicknell, Sutherland \& Neumayer}

\begin{document}

\title{The Kinematics and Ionization of Nuclear Gas Clouds in Centaurus~A }

\author{Geoffrey V. Bicknell\footnote{Geoff.Bicknell@anu.edu.au}, Ralph S. Sutherland\footnote{Ralph.Sutherland@anu.edu.au}}
\affil{Research School of Astronomy \& Astrophysics\\ Australian National University\\ Mt. Stromlo Observatory \\
Cotter Rd., Weston ACT, Australia 2611}

\and

\author{Nadine Neumayer\footnote{nadine.neumayer@universe-cluster.de}}
\affil{ European Southern Observatory\\ 
Karl-Schwarzschild Strasse 2\\
85748 Garching - Germany}

\begin{abstract}
Neumayer et al. established the existence of a blue-shifted cloud in the core of Centaurus A, within a few parsecs of the nucleus and close to the radio jet. We propose that the cloud has been impacted by the jet, and that it is in the foreground of the jet, accounting for its blue-shifted emission on the Southern side of the nucleus. We consider both shock excitation and photoionization models for the excitation of the cloud.  Shock models do not account for the [SiVI] and [CaVIII] emission line fluxes. However, X-ray observations indicate a source of ionizing photons in the core of Centaurus~A;  photoionization by the inferred flux incident on the cloud can account for the fluxes in these lines relative to Brackett-$\gamma$. The power-law slope of the ionizing continuum matches that inferred from synchrotron models of the X-rays. The logarithm  of the ionization parameter is -1.9, typical of that in Seyfert galaxies and consistent with the value proposed for dusty ionized plasmas \citep{dopita02a}. The model cloud density depends upon the Lorentz factor of the blazar and the inclination of our line of sight to the jet axis. For acute inclinations, the inferred density is consistent with expected cloud densities.   However, for moderate inclinations of the jet to the line of sight, high Lorentz factors imply cloud densities in excess of $10^5 \> \rm cm^{-3}$ and very low filling factors, suggesting that models of the gamma ray emission should incorporate jet Lorentz factors $\lesssim 5$.
\end{abstract}


\section{Introduction}
\label{s:intro}

\citet{neumayer07a} presented high resolution ($0.12^{\prime\prime}$) infrared spectral images of the central parsecs of Centaurus~A, which they utilized to derive important dynamical constraints on gas motions in this region, as well as the mass of the central black hole (see \citet{neumayer10a} for a review of the black hole mass measurements). A feature of the data, which has not yet been exploited is the existence of emission line gas clouds approximately aligned with the parsec-scale radio jet. Questions which naturally arise in this context are: What is the excitation mechanism for this gas and what is the cause of the $\sim 100 \> \rm  km \> s^{-1}$ velocities of this gas relative to the nucleus? Two candidates for the excitation mechanism are: (1) Shock waves driven by the jet into clouds adjacent to the jet and (2) Photoionization by either emission from an accretion disk or from the jet close to the back hole. Shock excitation could arise naturally in the process of entrainment of clouds into the jet or simply by jet deflection off these clouds. Considering the possibility of photoionization, \citet{lenain08a} modeled the radio through X-ray and gamma-ray emission from the nucleus of Centaurus~A in terms of either synchrotron plus synchrotron self-Compton (SSC) emission or two component synchrotron emission from a relativistic jet. In a subsequent paper they incorporated the newly discovered very high energy (VHE) gamma-ray emission from Centaurus A into a revised model in which the $1-10^5 \> \rm eV$ emission is due to synchrotron radiation. The significance of a synchrotron explanation for the $1-10^5 \> \rm eV$ emission  is that it may have a high enough flux (depending upon beaming effects) to be an important source of ionizing photons. If an SSC model is used for the $1-10^5 \> \rm eV$ flux, then the flux density of ionizing photons is much lower at around 10 eV (cf \citet{lenain08a}). If one or both of the jets in Centaurus~A are indeed responsible for the photoionization of the nuclear gas clouds then modelling of the emission may lead to interesting constraints on the Lorentz factor, which strongly affects the number density of ionizing photons through relativistic beaming.

In this paper, we summarize the observational data in \S~\ref{s:data}, consider both shock and photoionization models in \S~\ref{s:models} and discuss our results in \S~\ref{s:conclusions}. In an appendix, we derive expressions for the ionizing photon density incident on a nuclear cloud in terms of the observed flux density and the jet parameters relating to the beaming in both the observer and cloud directions.

\section{Observational data}
\label{s:data}

\begin{figure}[ht!]
\centering
\includegraphics[width=\textwidth]{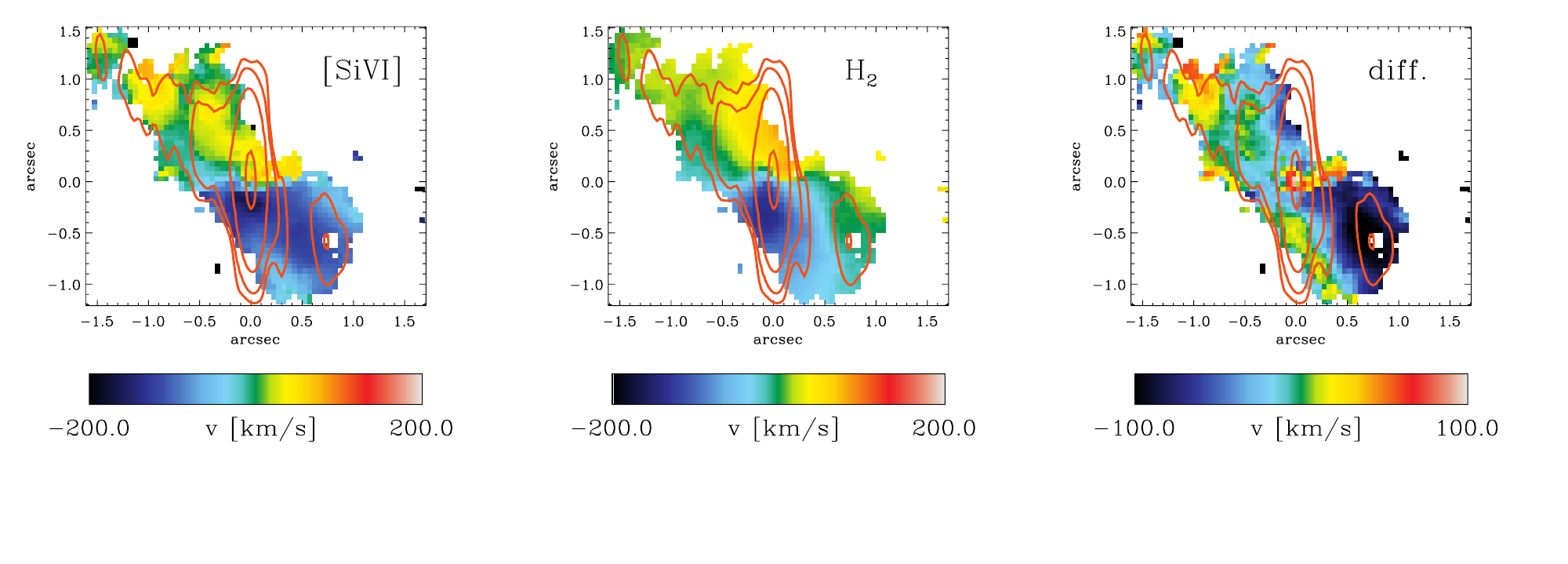}
\caption{This figure is reproduced from Figure~7 of \citet{neumayer07a}. The left and middle panels compare the [SiVI] and $\rm H_2$ velocity fields. The right hand panel shows the differenced velocity field of [SiVI] (i.e. $v([SiVI]) - v(\rm H_2)$ highlighting the blue-shifted cloud, coincident with the radio knot SJ1 \citep{hardcastle03a}.
(A color version of this figure is available in the online journal.) }
\label{f:Neumayer_Fig7}
\end{figure}

\begin{figure}[ht!]
\centering
\includegraphics[width=\textwidth]{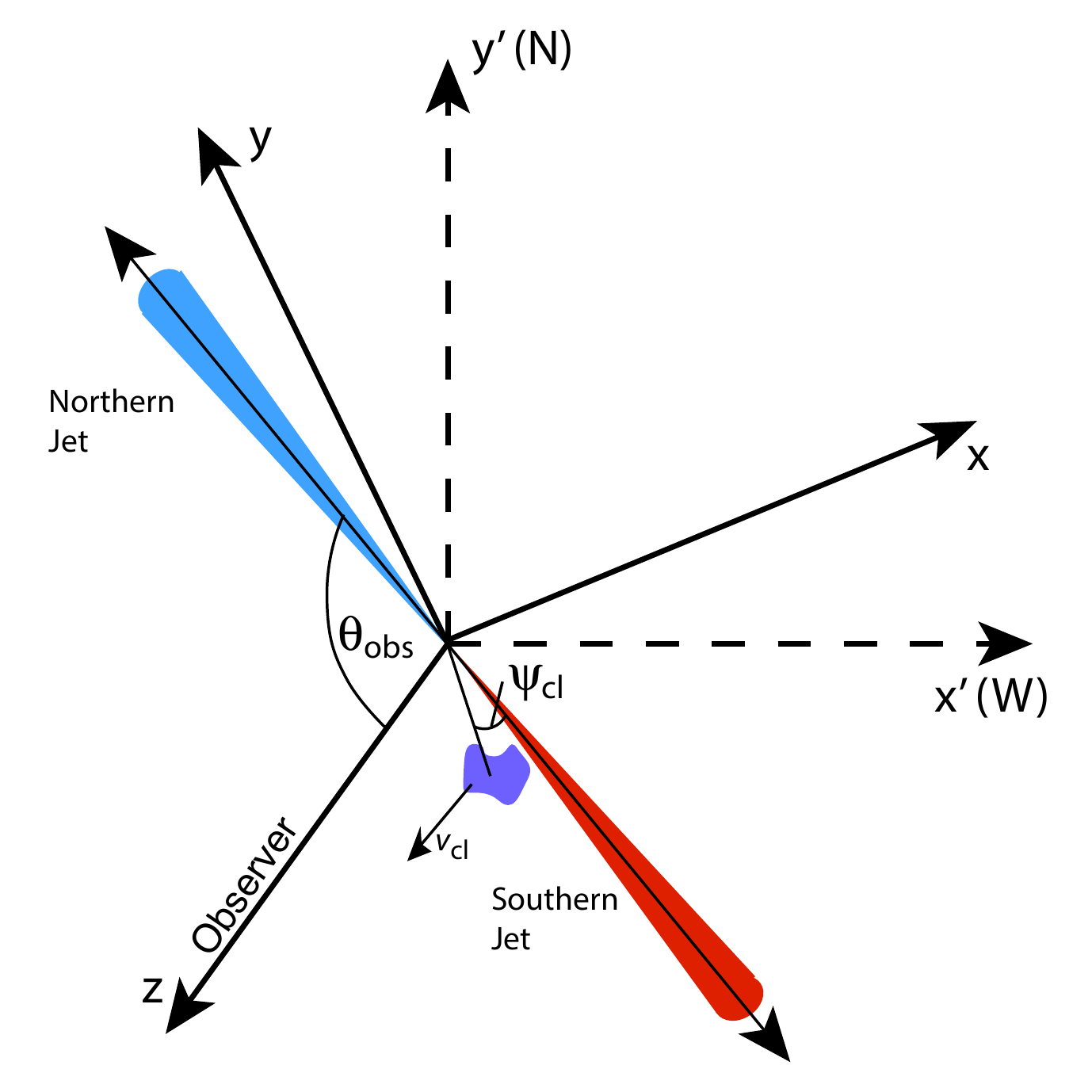}
\caption{The proposed geometry of the inner Centaurus~A jets, the radio knot SJ1 and the blue cloud. 
(A color version of this figure is available in the online journal.)}
\label{f:Geometry}
\end{figure}

Figure~\ref{f:Neumayer_Fig7}, reproduced from Figure~7 of \citet{neumayer07a} highlights the observational data, which are the focus of this paper. The left panel shows the [SiVI] emission spread along the inner few arcseconds of both jet and counter jet. The middle panel shows the $\rm H_2$ emission in the same region, whose velocity \citet{neumayer07a} show is indicative of disk rotation. Hence the differenced velocity field of [SiVI] in the right hand panel shows the non-rotational component of [SiVI]. The striking feature of the [SiVI] emission is the feature  displaced approximately south-west of the nucleus in right ascension and declination by $(0.7^{\prime\prime}, -0.6^{\prime\prime})$ with a blueshift of approximately $100 \> \rm km \> s^{-1}$. This feature is referred to as the ``blue cloud'' in the following. A blueshift for this cloud is at first surprising: Most analyses of Centaurus~A assume that the Southern jet is moving away from us, so that one expects a cloud, which may be entrained by the jet, should be redshifted. \citet{neumayer07a} noted an alternative possibility that the blue cloud could be falling in toward the nucleus. In this paper we adopt the view that the blue cloud is on the same side of the jet as us and that it is pushed to one side by the jet -- hence the blueshift. The geometrical configuration that we envisage is represented in Figure~\ref{f:Geometry}. 

Such an interpretation for the cloud velocity raises the question of why the cloud would not acquire a substantial forward momentum along the jet direction, leading to a redshift. Simulations of jets interacting with inhomogeneous media both in the form of disks \citet{sutherland07a} and spherically distributed clouds \citet{wagner11b,wagner11a} show that jet-cloud interactions are complex and are often affected by the backflow of radio plasma, either from the head of the radio source or from the downstream interaction of the jet with other clouds. The back flow impedes forward motion of the cloud so that its resultant motion is primarily perpendicular to the jet.

\section{EMISSION LINE MODELS}
\label{s:models}

In this section we discuss the two principal possible excitation mechanisms of gas in the nucleus of Centaurus~A: Shock excitation and photoionization. In doing so, we concentrate on the blue-shifted cloud  (see Figure~\ref{f:Neumayer_Fig7}) since this is separated in velocity from the rest of the gas in the nuclear region and offers the prospect of being described by a well-defined set of parameters of number density and photon density. However, we also consider the integrated emission from the nucleus.

The K-band infrared spectra do not offer a large number of emission lines for comparison with the models. The principal lines that are evident in the spectra are those of [SiVI] at $1.9602 \> \mu \rm m$, Br$\gamma$ at $2.1656 \> \mu \rm m$ and [CaVIII] at $2.322 \> \mu \rm m$ so that the models are not  uniquely constrained by the data. However, [CaVIII] is an important discriminant of shock-excitation and photoionization and it is possible to derive reasonable conclusions on the relative merits of these two mechanisms and also to estimate the cloud densities.

\subsection{Shock Models}  

\begin{figure}[ht!]
\centering
\includegraphics[width=\textwidth]{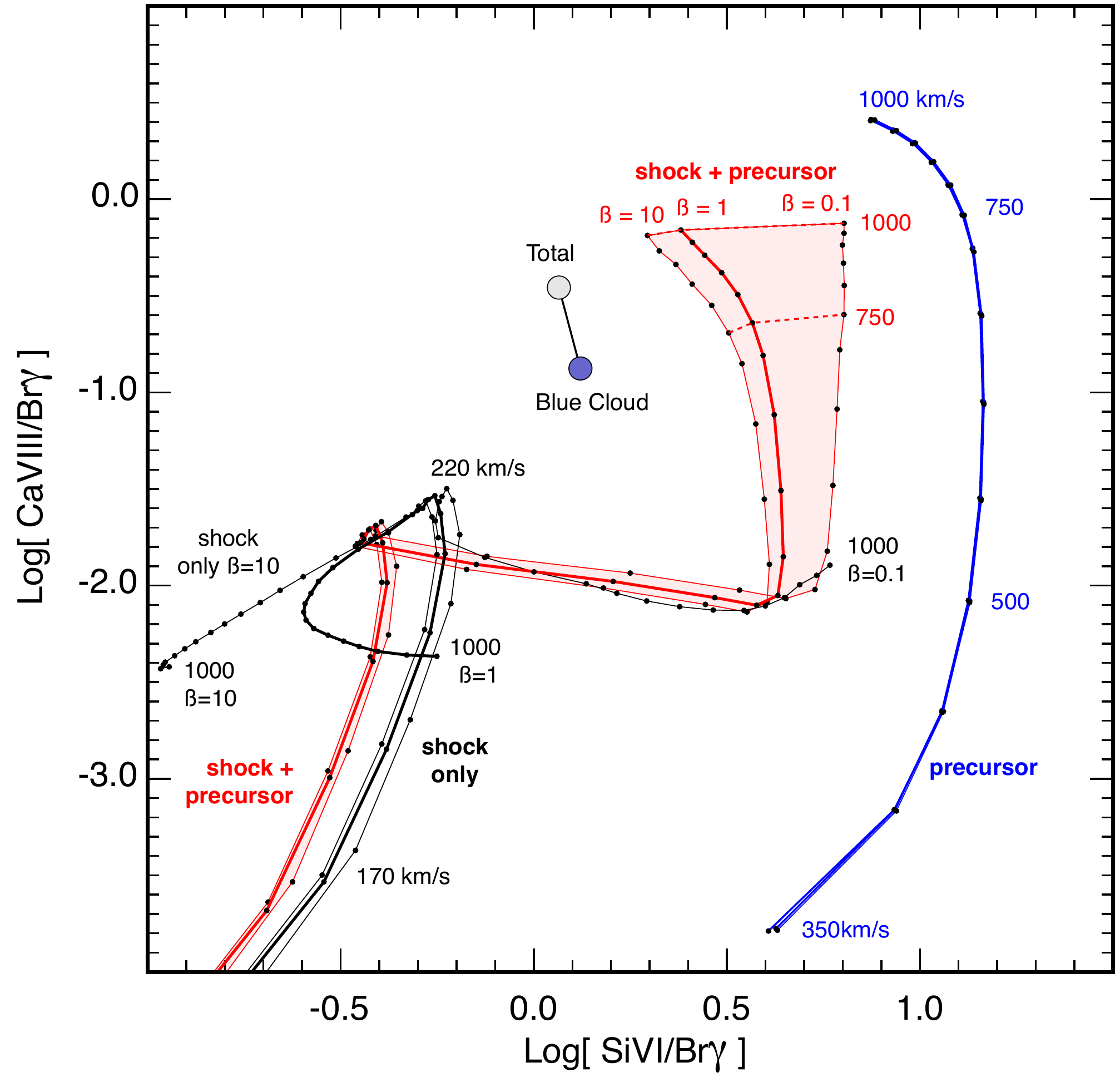}
\caption{Shock models of the ratio of [CaVIII] to Br$\gamma$ plotted against the ratio of [SiVI] to Br$\gamma$ for the nuclear region of Centaurus A. Combined shock + precursor models as well as shock-only models and precursor-only models are shown together. The magnetic field is parameterized by $\beta$, the ratio of thermal pressure to magnetic pressure in the pre-shock gas. Note that the parameter $\beta$ does not affect the precursor. (A color version of this figure is available in the on-line version.)}
\label{f:shock_models}
\end{figure}

Since we are suggesting that the velocity of the blue cloud is the result of being pushed to one side by the passage of the jet it is natural to investigate whether the emission from the cloud could be the result of shock excitation.

The emission from an astrophysical shock wave comprises contributions from the shocked gas and the precursor region photoionized by radiation from the shock. Using the MAPPINGS~III emission line code, we have computed a comprehensive grid of shock + precursor models for velocities between 160 and 1000 
$\rm km \> s^{-1}$. The results for [CaVIII] and [SiVI] are shown in Figure~\ref{f:shock_models} wherein the shock and precursor regions are shown both separately and combined. The effect of varying the pre-shock magnetic field is also shown in Figure~\ref{f:shock_models}.

The compelling message from the shock models is that there is no combination of shock velocity and magnetic field, which adequately reproduces both the observed [CaVIII]/Br$\gamma$ and [SiVI]/Br$\gamma$ line ratios. The adjacency of gas to the jet combined with its velocity offsets with respect to the core indicates that its kinematics may be affected by the jet. However, it does not appear that shock waves associated with a jet-cloud interaction dominate the emission. 

\subsection{Photoionization Models}
\label{s:photo}

We now consider photoionization models for the emission from the nuclear region of Centaurus~A. The point of view that we take here, is inspired by that proposed by \citet{pedlar85a} and \citet{taylor92a} for Seyfert galaxies. Clouds of moderate density are impacted by the radio jets producing higher density radiatively shocked regions. The shocked gas is photoionized by the nuclear continuum, producing the observed emission line spectrum.

\subsubsection{Photoionizing Flux}
\label{s:photoionizing_flux}

One possible source of the ionizing continuum is the nuclear sub-parsec-scale jet. In order to assess the implications of such a source, which may be relativistically beamed,  we need to determine the flux of ionizing radiation incident on the cloud given the radiation that we observe along a line of sight, which is inclined at a different angle to the velocity of the radiating plasma.

The high energy emission of the \emph{Northern} jet from hard X-rays through to very high energy (Tev) gamma rays has been modeled by \citet{lenain09a} in terms of a Synchrotron-Self-Compton (SSC) model. This was an update of an earlier model \citep{lenain08a} in which both synchrotron and inverse Compton models were considered for the X-ray emission alone. However, the detection of VHE $\gamma$-ray emission from Centaurus~A \citep{aharonian09b} has made the synchrotron interpretation more appealing. This is important since the inverse Compton model has a much lower ionizing flux. In the model that we develop here, we assume that the base of the \emph{Southern} jet has a similar spectrum to the Northern jet and that this provides the flux of photoionizing radiation.

Figure~1 in \citet{lenain09a} shows that the continuum region that is of interest here, namely the ionizing continuum with energies in excess of approximately 10~eV in the rest frame of the irradiated gas, is only constrained by the observational data in the range $\sim 1-100 \> \rm keV$. Hence we rely on the \emph{modeled} spectrum in the region 
$\sim 1 - 10^5 \> \rm eV$. The model spectral energy distribution (SED) between approximately 1~eV and 100~keV (the power-law synchrotron component indicated on their Figure~1) has a flux spectral index of 
$\alpha \approx 0.39$ ($F_\nu \propto \nu^{-\alpha}$). This is slightly flatter than usual (say $\sim 0.6-0.7$) for a non-thermal spectrum. Nevertheless, values of 0.4 and 0.3 have been inferred for the synchrotron component of the blazar emission in MKN~501 by \citet{bicknell01b} and \citet{konopelko03a} respectively.

\citet{lenain09a} present three different models. In the following we use the model which most accurately reproduces the X-ray data, the thin line in their Figure~1.

\citet{lenain09a} infer a Lorentz factor of 15 for the Centaurus~A jet. Lorentz factors of this magnitude are controversial in view of the fact that Lorentz factors estimated on the parsec scale from radio observations are always lower than this \citep[e.g.][]{piner08a} and Centaurus~A is no exception: The pattern speeds of knots in the Centaurus~A jets are of order 0.1~c \citep{tingay98a} -- much less than the superluminal pattern speeds seen in many quasars, although Tingay et al. note that there is evidence for a faster underlying flow. Estimated high Lorentz factors in blazars may point to significant deceleration between the sub-parsec and parsec scales or to localized fast-moving regions of the flow and these issues are currently unresolved. However, the inference of high Lorentz factors and the observation of superluminal motions in AGN in general, means that we need to take account of relativistic beaming. As noted above, in Centaurus~A the radiation field observed by us should be different from the radiation field intercepted by the blue-shifted cloud and it is important to allow for this in the photoionization models.

The various relationships required to relate the observed flux density to the ionizing photon density at the cloud are derived in the appendix. We summarize the main results here:

Our model is summarized in Figure~\ref{f:Geometry}. We envisage the inner jet, within say about 100 gravitational radii ($\sim 20$~mpc) as the source of ionizing photons (the inner jet); let $D_{\rm cl}$ be the distance of the cloud from the source; let $D_{\rm A}$ be the angular diameter distance of Centaurus~A; let the Doppler factors of the radiation received by observer and cloud respectively be $\delta_{\rm obs}$ and $\delta_{\rm cl}$ and let $z\approx 0.0018$ \citep{graham78a} be the redshift of Centaurus~A. If $\nu$ is the frequency of a photon intercepted by the cloud, it is emitted by the jet with a rest frame frequency $\delta_{\rm cl}^{-1} \nu$ and a photon emitted with this frequency reaches the observer with a frequency $\nu_{\rm obs} = \delta_{\rm obs} (1+z)^{-1} \delta_{\rm cl}^{-1} \nu$. Let $F_{\rm obs}(\nu_{\rm obs})$ be the observed flux density of the inner jet as a function of the observing frequency $\nu_{\rm obs}$. 

With these definitions, the photon density per unit frequency at the cloud (derived in the Appendix) is
\begin{equation}
n_{\rm ph}(\nu) =  \frac {1}{c} \, \left( \frac {D_{\rm A}} {D_{\rm cl}}\right)^2 \, 
\left[\frac {\delta_{\rm cl}(1+z) }{\delta_{\rm obs}} \right]^3 \,
\, \frac {F_{\rm obs} (\delta_{\rm obs} (1+z)^{-1} \delta_{\rm cl}^{-1} \nu)}{h \nu}
\label{e:n_ph_nu}
\end{equation}

Let $\nu_0 \doteq 3.28 \times 10^{15} \> \rm Hz$ be the frequency corresponding to the Rydberg limit. The total number density of ionizing photons at the cloud (also derived in the appendix) is 
\begin{eqnarray}
n_{\rm ph} &=& \int_{\nu_0}^\infty n_{\rm ph}(\nu) \> d \nu
\nonumber \\
&=& \frac {1}{c} \, \left( \frac {D_A}{D_{\rm cl}} \right)^2 \, 
\left[ \frac {\delta_{\rm cl} (1+z) }{\delta_{\rm obs}} \right]^3  \>
\int_{\delta_{\rm obs} (1+z)^{-1} \delta_{\rm cl}^{-1} \nu_0}^\infty \, \frac {F_{\rm obs}(\nu_{\rm obs})}{h\nu_{\rm obs}} \> d\nu_{\rm obs}
\label{N_ph_final}
\end{eqnarray}

Between about $10^{14.5}$ and an upper frequency $\nu_u \approx 10^{19.3} \> \rm Hz$ the modeled spectrum may be described as a power-law $F(\nu_{\rm obs}) \approx F(\nu_0) (\nu_{\rm obs}/\nu_0)^{-\alpha}$ with $F(\nu_0) \approx 1.35 \times 10^{-27} \> \rm ergs \> s^{-1} \> Hz^{-1}$ and $\alpha \approx 0.39$. Hence, 
\begin{equation}
n_{\rm ph} \approx \frac {F_{\rm obs}(\nu_0)}{\alpha c h} \left( \frac {D_A}{D_{\rm cl}}\right)^2 
\, \left[ \frac {\delta_{\rm cl} (1+z)}{\delta_{\rm obs}} \right]^{3+\alpha} \,
\left\{ 1 - \left[\frac{\delta_{\rm cl}(1+z)}{\delta_{\rm obs}}\frac{\nu_u}{\nu_0} \right]^{-\alpha} \right\}
\label{e:n_ph}
\end{equation}
Since $\nu_u \gg \nu_0$, the contribution of the second term in brackets is minor.

Let the inclination of our line of sight to the jet be $\theta_{\rm obs}$, the angle between the direction of a ray from the core through the center of the cloud and the jet be $\psi_{\rm cl}$ and the (projected) angular displacement of the cloud from the core be $\xi_{\rm cl}$, the geometry of the model (see Figure~\ref{f:Geometry}) implies that the ratio of angular diameter distance to cloud distance is given by
\begin{equation}
\frac{D_A}{D_{\rm cl}} = \frac {\sin (\theta_{\rm obs}+\psi_{\rm cl})}{\xi_{\rm cl}}
\label{e:da_dcl}
\end{equation}
(This equation is also derived in the Appendix.) Furthermore, the ratio of Doppler factors is given by:
\begin{equation}
\frac {\delta_{\rm cl}}{\delta_{\rm obs}} = \frac {1 - \beta \cos \theta_{\rm obs}}{1 - \beta\cos \psi_{\rm cl}}
\end{equation}
As expected, the expression for the ionizing photon density has a strong dependence on the ratio of Doppler factors, when 
$\psi_{\rm cl}$ is small. We estimate $\psi_{\rm cl}$ assuming that the cloud is adjacent to the jet and that is depth along the line of sight is the same as its transverse diameter $\approx 0.35^{\prime\prime}$. The expression for $\psi_{\rm cl}$ is derived in the Appendix (see equation~\ref{e:psi_cl}).

\subsubsection{Isochoric Photoionization Models}

\begin{figure}
\includegraphics[width=\textwidth]{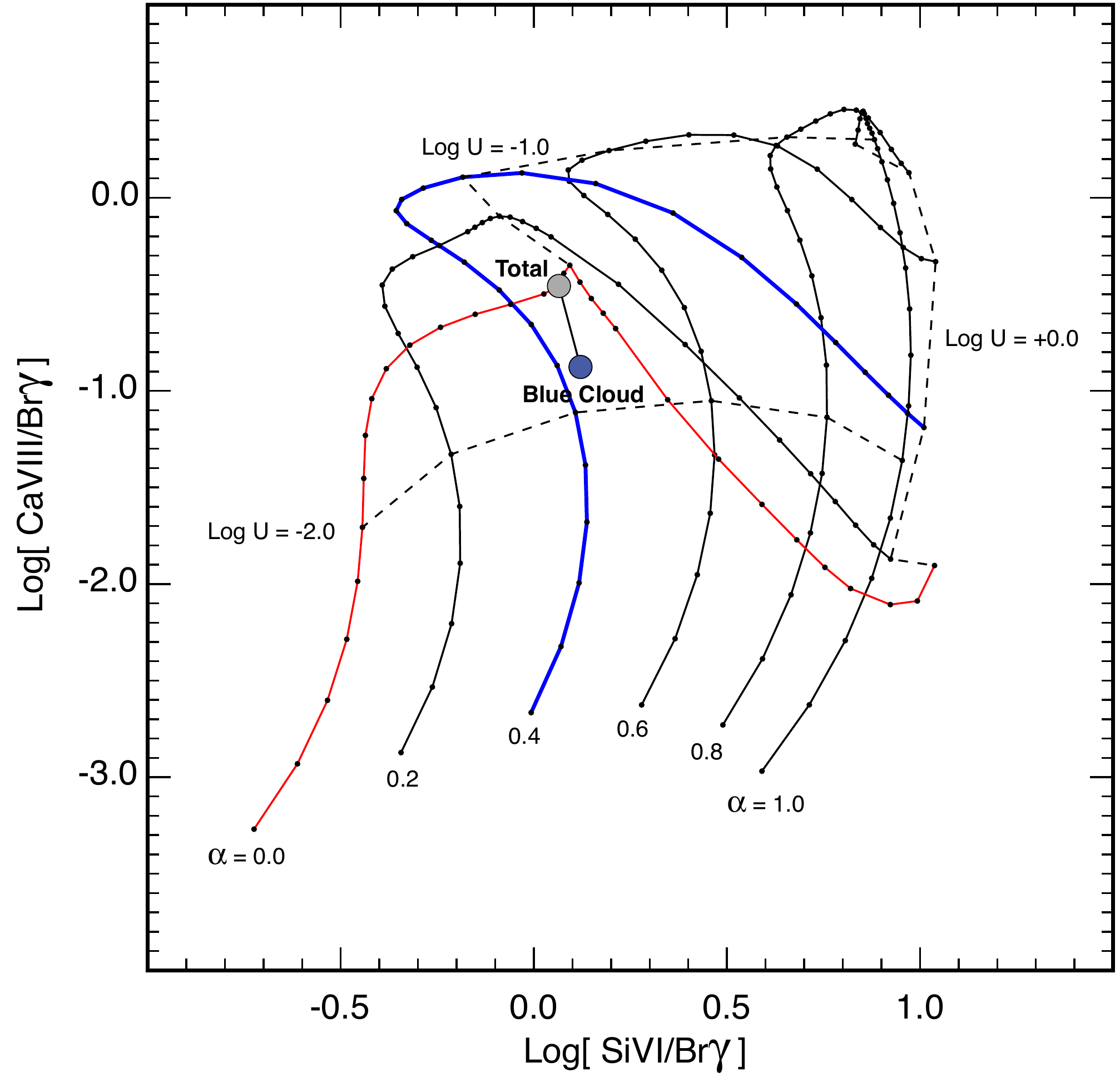}
\caption{The line ratios [CaVIII]/Br$\gamma$ and [SiVI]/Br$\gamma$ resulting from a grid of MAPPINGS isochoric, photoionization models for solar metallicity. The spectral indices are marked on the different curves and the dots on the curves represent increments of 0.1 in the logarithm of the ionization parameter. Data points corresponding to (a) the blue cloud and (b) the total nuclear emission are also marked.}
\label{f:photo}
\end{figure}

In framing models for cloud photoionization one question to resolve is whether an isobaric or isochoric model is more appropriate.
Here, an isobaric model is relevant if the expansion time scale for the gas heated by photoionization is less than the time scale for cloud shredding resulting from the jet-cloud interaction. The photoionized cloud would expand at approximately the sound speed corresponding to a temperature of $10^4 \> \rm K$, that is, about $12 \> \rm km \> s^{-1}$. If we take the anomalous velocity of the blue cloud $\sim 100 \> \rm  km \> s^{-1}$ as indicative of shocks driven into it by the jet, then the shock-shredding time scale $\sim 18 \> {\rm pc} / 100 \> \rm km \> s^{-1} \approx 1.8 \times 10^5 \> \rm yrs$ is approximately a factor of eight shorter than the expansion time scale. Hence, we mainly consider isochoric, constant density, models in the following.

We have calculated a grid of models for solar metallicity using the solar abundance scale of \citet{grevesse10a}. The models are parameterized by the ionization parameter, $U$, the ratio of the ionizing photon density, $n_{\rm ph}$ to the particle density $n$ ($U = n_{\rm ph}/ n$) and the power-law index, $\alpha$ of the photoionizing flux density ($F_\nu \propto \nu^{-\alpha}$). The models are ionization bounded and are for a density $n=10^4 \> \rm cm^{-3}$. For this fiducial density, the extent of the ionization bounded region $\approx 10^{19.5} \rm cm$ and fits comfortably within the projected major axis $\approx 10^{20.7} \> \rm cm$  of the blue cloud. Below, we consider lower filling factors, involving higher densities. However, the model line ratios remain independent of density for all the densities considered here.

Figure~\ref{f:photo} shows the results of the grid of models for the line ratios [SiVI]/Br$\gamma$ and [CaVIII]/Br$\gamma$. The 
best-fit model is represented by $(\log U,\alpha) \approx (-1.9, 0.42)$. The best-fit value of $\alpha$ is in practice indistinguishable from the value inferred from the synchrotron emission fit to the X-ray data, strongly supporting the notion of photoionization by jet emission.

We note that the best fit ionization parameter of $\log U \sim -2.0$ is in rough agreement with 
the self-limiting radiation pressure dominated dusty AGN nebula models of \citet{dopita02a,groves04a,groves04b,groves04c}.
In future work it may be fruitful to explore the possibility of such dusty models in the present context.
However, with a scarcity of diagnostic lines such as [OIII]~$\lambda5007$, and a lack of 
constraints on the properties of the dust, and any depletion factors that may be present in the 
blue cloud (affecting the abundances of both Silicon and Calcium), additional model parameters are not justified by the present spectral data.

If more data on the cloud composition, dust properties and more spectral line measurements were 
available, then radiation pressure dominated dusty models may improve the models and place
better constraints on the Br$\gamma$ emission for example. This would affect the filling factor of the cloud (see below).

In our isochoric models, the ratios of [SiVI] and [CaVIII] to Br$\gamma$ are primarily determined by the slope of the ionizing continuum and the level of the emission is determined by the density. Hence, we can estimate the cloud number density, $n$ from the ionization parameter, $U$, and the ionizing photon density, $n_{\rm ph}$ (using $n = U^{-1} \, n_{\rm ph}$). Since the ionizing photon density depends upon the observed continuum, the angle of inclination of the radio jet and its Lorentz factor, through equation~(\ref{e:n_ph}), we can assess the effect of these quantities on the inferred density of the cloud. Note that our estimates of the number density would not be substantially revised by dusty models since such models predict a very similar ionization parameter to what we have inferred here.

\subsubsection{Filling Factor.}

The filling factor of the Hydrogen-line emitting region of the cloud can be estimated from the Br$\gamma$ luminosity, 
$\rm L(Br\gamma)$. Let $\alpha(\rm Br\gamma)$ be the effective Case B recombination coefficient for Br$\gamma$, $f$ the volume filling factor, $n_e$ and $n_H$ the electron and Hydrogen number densities and $\rm V_{\rm cl}$ the cloud volume, then
\begin{equation}
L({\rm Br}\gamma) = \alpha({\rm Br}\gamma) \, f \, n_e n_H \, V_{\rm cl}.
\label{e:f}
\end{equation}
We estimate a cloud volume 
\begin{equation}
V_{\rm cl}  \approx  D_A^3 \times (\pi/6) \> \theta_{\rm maj} \theta_{\rm min}^2 / \sin \theta_{\rm obs}
\label{e:volume}
\end{equation}
where $\theta_{\rm maj} \approx 1.1^{\prime\prime}$ and $\theta_{\rm min} \approx 0.7^{\prime\prime}$ are the angular sizes of the major and minor axes of the cloud respectively; we assume that the depth of the cloud is similar to the length of the minor axis. 

In using Equations~(\ref{e:f}) and (\ref{e:volume}) to estimate the filling factor, we use the measured value of the Br$\gamma$ flux of $2.42 \times 10^{-16} \rm ergs \> s^{-1} \> cm^{-2}$ and adopt a value of $2.67 \times 10^{-27}$ for the Br$\gamma$ emission coefficient, appropriate for the average temperature of $12,500 \> \rm K$ obtained in the photoionization models. Note that the estimate for the filling factor depends upon the assumption that the cloud depth is similar to its width. If the  depth were an order of magnitude smaller, for example as a result of ablation by the jet, then the filling factor would increase by an order of magnitude. 

\begin{figure}[ht!]
\plottwo{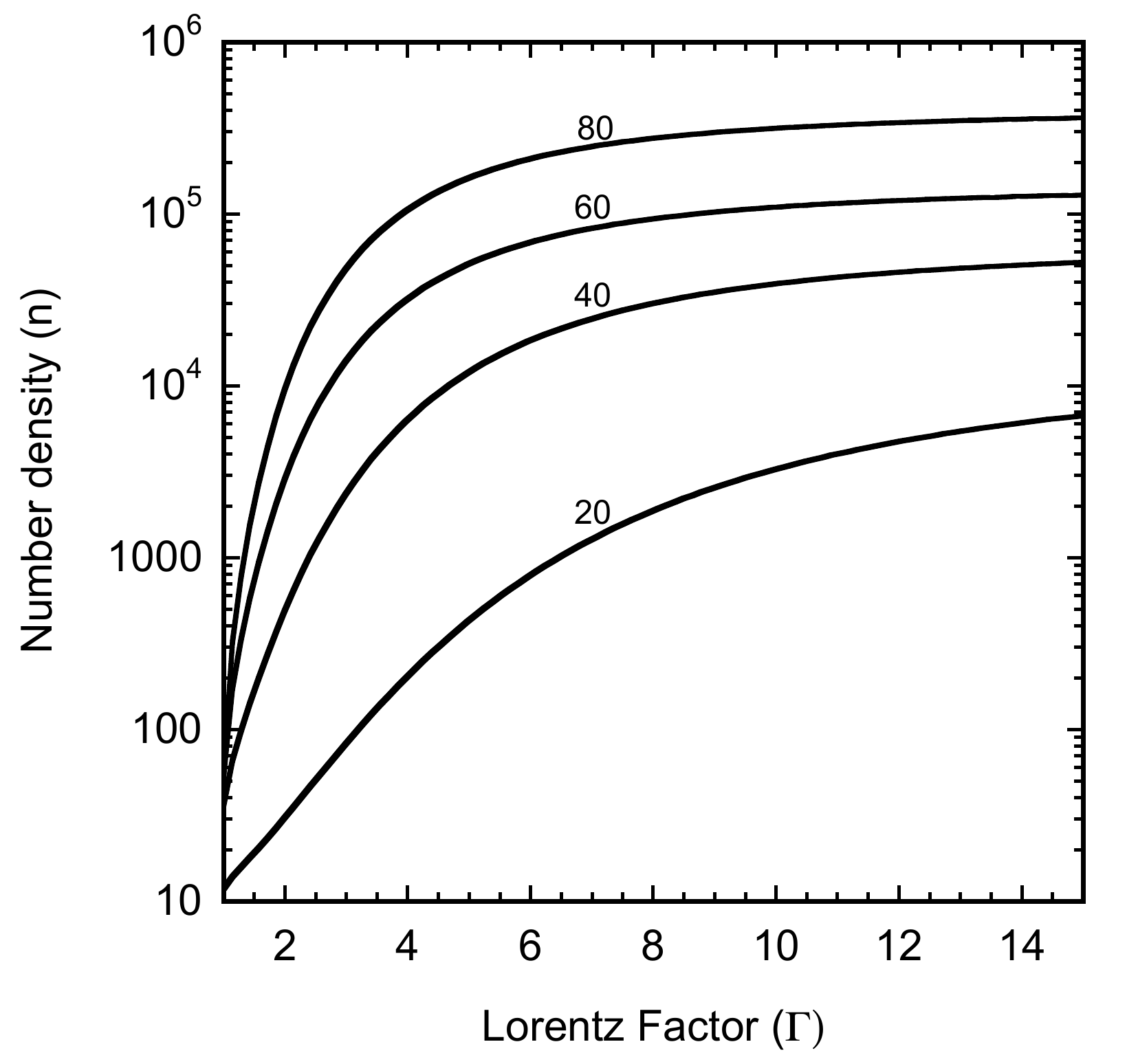}{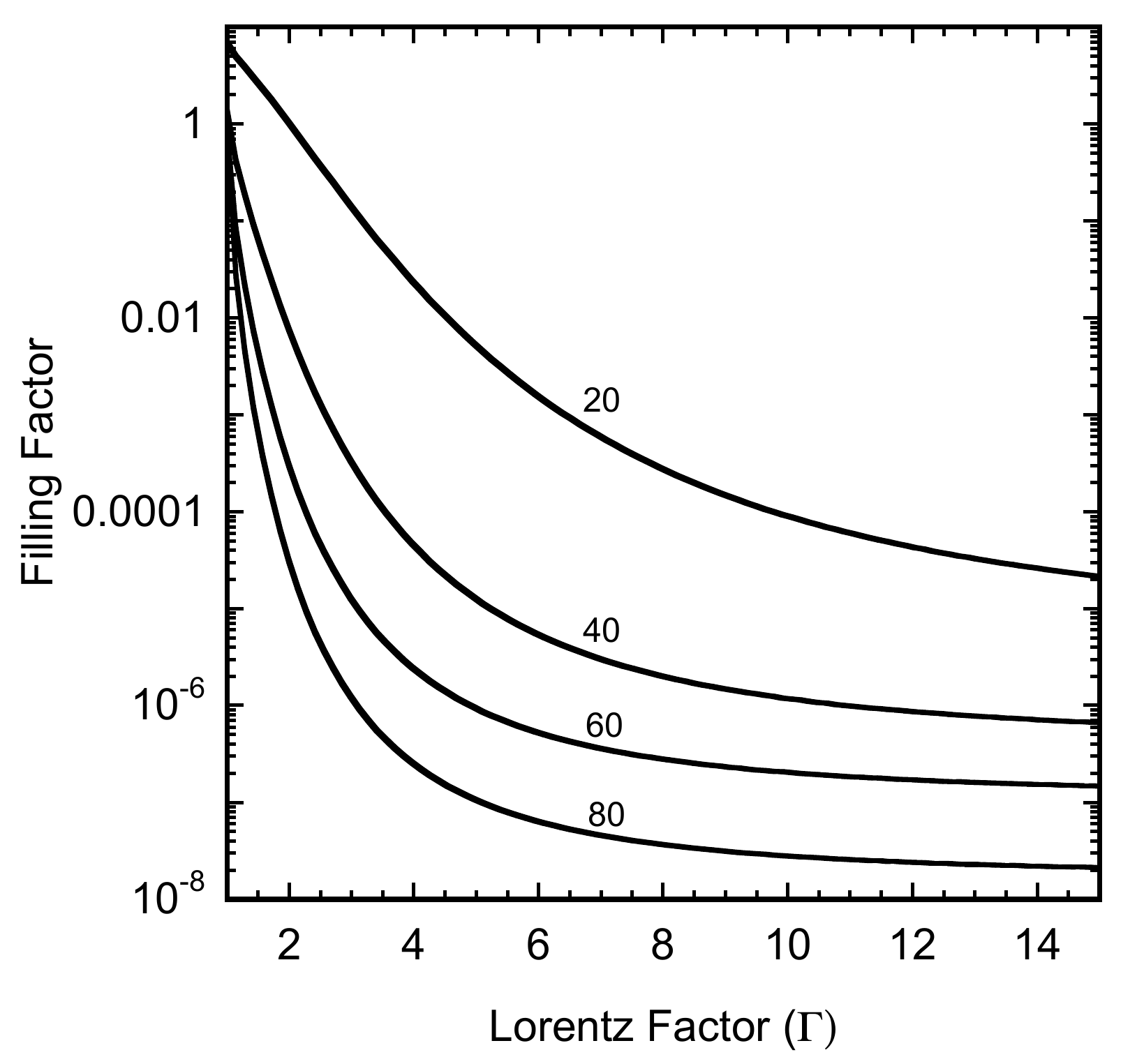}
\caption{Left panel: Estimate of the cloud density from the ionization parameter and the ionizing photon density, which is a function of the jet Lorentz factor and the inclination of the jet to the line of sight. Right panel: Estimate of the filling factor from the Br$\gamma$ luminosity.}
\label{f:n_and_ff}
\end{figure}

Our estimates of the density of the photoionized gas and the related filling factor both as a function of Lorentz factor and 
$\theta_{\rm obs}$ are presented in the left and right panels of Figure~\ref{f:n_and_ff}.  For all $\theta_{\rm obs}$, the density increases quite rapidly with Lorentz factor. For $\theta_{\rm obs} = 40^\circ, 60^\circ$ and $80^\circ$, the density increases to values in excess of $10^4 \> \rm cm^{-3}$. In keeping with this, the filling factor decreases quite rapidly with Lorentz factor for all inclination angles but especially for $\theta_{\rm obs} \ge 40^\circ$. The strong dependence of the density and filling factor on Lorentz factor results from the effects of beaming and the high inferred photon density at the cloud when the line of sight to the observer is greater than approximately $30^\circ$ to the jet axis (see equation~(\ref{e:n_ph})). That is, for moderate values of 
the angle of the observer to the line of sight, the observer sees a considerably lower intensity than that which is incident on the cloud.

Various inclination angles have been proposed for the Centaurus~A jet but most are within the range $50^\circ \le \theta_{\rm obs} \le 80^\circ$ proposed by \citet{tingay98a}. The exception is \citet{hardcastle03b} who estimate an angle of inclination of approximately $15^\circ$ \emph{if the inner Very Large Array (VLA) jets are intrinsically symmetric}. However, they note the inconsistency of this estimate with other estimates of inclination, leaving open the question of the asymmetry of the inner VLA structure. 

What is the evidence for gas with densities in excess of $10^3 \> \rm cm^{-3}$? Using a spherical free-free absorption model for the very long baseline interferometry (VLBI) jets, \citet{tingay01a} derived a number density $\sim 9 \times 10^4 \, T_4^{1.35}$ where the temperature of the absorbing gas is $10^4 \, T_4 \> \rm K$ and the radius of the free-free absorbing sphere is 0.016~pc. \citet{wild97a} determined gas densities $\approx 1-3 \times 10^4 \> \rm cm^{-3}$ from CO observations of the Centaurus~A dust lane, at a spatial resolution of $44^{\prime\prime}$. Subsequent work \citep{wild00a} using HCN and CS observations at a resolution of $54^{\prime\prime}$ showed that approximately 10\% of the molecular gas in the nuclear region has densities in excess of $10^4 \> \rm cm^{-3}$. Thus, there is good evidence for dense gas clouds in the nuclear region of Centaurus~A, associated with the dust lane. Future ALMA observations should provide more information on the spatial distribution of these clouds. However, we do not expect the cloud densities at a few parsecs to be higher than the $10^5 \> \rm cm^{-3}$ inferred by \citet{tingay01a} near the core so that we adopt a conservative upper estimate on the density of $10^5 \> \rm cm^{-3}$. 

It is only for inclinations $\la 50^\circ$ that the constraint $n < 10^5 \> \rm cm^{-3}$ is realized for all Lorentz factors up to the estimate of $\Gamma=15$ by \citet{lenain09a}. This density constraint is consistent with the \citet{lenain09a} models in which a low inclination angle, $\theta_{\rm obs}=25^\circ$, is assumed but not with the range of inclinations discussed above.

Let us now consider the dependence of filling factor on Lorentz factor. While describing the density structure of a photoionized region using this parameter is not ideal, we adopt if here for the purpose of approximate comparisons. In the narrow line region of active galaxies the value inferred for the filling factor is typically $\sim 10^{-4} - 10^{-3}$. \citet{pedlar85a} and \citet{taylor92a} explained this low value in terms of the dynamics of radio plasma driven radiative shocks, which produce high density, but low volume filling factor, shocked regions. The high density regions are photoionized by the nuclear radiation and the Pedlar et al. and Dyson et al. models explain why photoionization models require a low filling factor. Pedlar et al. and Dyson et al. formulated their models to explain the observed properties of Seyfert galaxies, and the physical situation envisaged here is similar to their Seyfert model: Gas shocked by the jet is photoionized by the central source producing regions of photoionized gas with a low filling factor. Hence a comparison of the filling factors in the narrow line regions of Seyfert galaxies and those deduced here for Centaurus A is useful. We therefore adopt a conservative \emph{lower} limit on the filling factor of approximately $10^{-4}$.

When we examine the dependence of filling factor on Lorentz factor (right panel of Figure~\ref{f:n_and_ff}) we see that the inferred filling factor decreases rapidly with Lorentz factor. This results from the increase of the beamed photon density and atomic density with Lorentz factor. The effect is highlighted for moderate inclinations. For acute inclinations the value of the filling factor is acceptable up to $\Gamma \approx 10$. However, for moderate inclinations, filling factors $\la 10^{-4}$ are only possible for Lorentz factors less than a few.

\section{Conclusions and discussion}
\label{s:conclusions}

The discovery of blueshifted line-emitting gas on the side of the nucleus that we would normally associate with redshifted outflow has led us to consider detailed models for the location, kinematics and excitation of this gas.

The features of the model that we have investigated in detail include the following: 
\begin{enumerate}
\item  The blue-shifted cloud has been impacted by the Southern jet and is located close to the jet and between the jet and us. 
\item  The gas has been shocked by the jet producing dense sub-clouds/filaments of gas but the primary excitation mechanism for the line emission is not shock excitation but photoionization by ultraviolet and X-ray emission from the core of Centaurus~A. 
\item The X-ray emission from the core combined with the observations and models of the high energy gamma-rays \citep{lenain09a} suggest that the ionizing flux originates from the Southern jet. We have assumed that the Southern jet flux is similar to that of the Northern jet. 
\item Photoionization models of the emission from the blue cloud require a power-law slope almost identical to the \citet{lenain09a} models of the  UV--X-ray emission from the core of Centaurus~A, in order to reproduce the [CaVIII] and [SiVI] fluxes relative to Br$\gamma$. 
\item The power-law slope is characteristic of the high-energy slopes of core jets in blazars.
\item The last two results indicate very strongly that ionizing emission from a relativistic jet in Centaurus~A is responsible for the excitation of not only the blue cloud but other nuclear line emission in Centaurus~A.
\item Consideration of both the number density and volume filling factor implied by the photoionization models and the flux of Br$\gamma$, strongly suggest, that the Lorentz factor of the jets in Centaurus A is lower than indicated by the high energy models. This is not surprising since it has been an unresolved problem for some time that blazar models usually imply higher Lorentz factors than those inferred from VLBI observations on the parsec scale.
\end{enumerate}

As we have seen in \S~\ref{s:photo} precise conclusions concerning the Lorentz factor of the jet in Centaurus~A (and consequently the viability of models for the high energy emission) depend strongly on our viewing angle and the interpretation of the milliarcsecond and arcsecond radio data. Broadly speaking, if the inclination of our line of sight to the jet axis is as acute as the interpretation of the arcsecond radio data may suggest \citep{hardcastle03b} then our photoionization models are consistent with a reasonably high Lorentz factor (for both jets). If, on the other hand, the inclination is in the range of $50^\circ - 80^\circ$ inferred from the millarcsecond radio data \citep{tingay98a}, then low to moderate Lorentz factors ($\Gamma \lesssim 5$) are implied . The is consistent with the lack of evidence for very high Lorentz factors in the form of superluminal proper motions \citep{tingay98a} such as are found in quasar jets for example \citep{kellermann04a}.

The physical distance, $D_{\rm cl}$  of the blue-shifted cloud from the nucleus is determined by the model parameters, specifically $\theta_{\rm obs}$, the angle between the observer and the Northern jet direction, and $\psi_{\rm cl}$, the angle between the direction of the southern jet and the direction of the cloud from the nucleus, together with the projected distance from the nucleus $\approx 16.5 \> \rm pc$ corresponding to an angular separation of approximately 1 arcsec. (See 
Figure~\ref{f:Geometry} and Equation~\ref{D_cl}.) For the most acute angle of inclination we have considered, $\theta_{\rm obs} = 20^\circ$, $D_{\rm cl} \approx 36 \> \rm pc$. For the range of angles referred to in the above paragraph $50^\circ < \theta_{\rm obs} < 80^\circ$ the distance of the cloud from the core varies very little: $18 \> {\rm pc} >  D_{\rm cl} > 17 \> \rm pc$. 
Note however, it is only in the case of the photoionization models that these parameters are relevant. The main parameter in the shock models is the shock velocity.

Our conclusion that the blue cloud is photoionized by beamed emission from the core invites the question as to the relevance of photoionization for other clouds on the kiloparsec scales in the interstellar medium of Centaurus~A \citep{blanco75a,peterson75a,graham81a,morganti91a}. In particular \citet{morganti91a} argued that the region referred to as  the inner filaments are photoionized by beamed radiation from the core. On the other hand, \citet{sutherland93b}, presented a model in which the filaments are locally shock-excited through interaction with the radio plasma.
More recently \citet{sharp10a} have found an ionization cone in the inner galaxy of Centaurus~A aligned with the radio jet as well as other nearby emission line regions, which are excited by star formation. In the light of the calculations presented in this paper, we can consider some examples based on typical parameters which indicate the conditions under which  photoionization may be feasible for the different regions well outside the core. Note that these regions are all to the North of the nucleus so that we do not need to assume the equivalence of Southern and Northern jet ionizing fluxes.

\begin{table}[ht!]
\centering
\begin{tabular}{c c c c c c }
\hline
Region     & $\xi_{\rm cl}$ & $\Gamma$ & $\psi_{\rm cl}$   & $n$ & $U$ \\
           & arcsec         &          &    degrees        & $\rm cm^{-3}$ & \\
\hline
Ionization & 25           &    2     &     17.5            & 100 & $4.1 \times 10^{-4}$ \\
Cone       & 25           &    4     &     17.5            & 100 & $4.9 \times 10^{-3}$ \\
           & 25           &    5     &     17.5            & 100 & $7.9 \times 10^{-3}$ \\
\hline
Inner      & 550          &    2     &     3.5             & 30  & $2.2 \times 10^{-5}$ \\
Filaments  & 550          &    4     &     3.5             & 30  & $1.8 \times 10^{-3}$ \\
           & 550          &    5     &     3.5             & 30  & $6.8 \times 10^{-3}$ \\
\hline
Outer      & 910          &    2     &     10.4            & 10  & $5.9 \times 10^{-6}$ \\
Filaments  & 910          &    4     &     10.4            & 10  & $2.0 \times 10^{-4}$ \\
           & 910          &    5     &     10.4            & 10  & $4.8 \times 10^{-4}$ \\
\hline
\end{tabular}
\caption{Estimates of Ionization Parameter in Different Regions of Centaurus A}
\label{t:U}
\end{table}

In Table~\ref{t:U} we present some indicative calculations of ionization parameters for these different regions for various values of jet Lorentz factor, nominal values of number density and a jet inclination of $60^\circ$. The angle between jet and cloud directions were estimated as follows. For the ionization cone \citet{sharp10a} estimated a half-opening angle of $20^\circ$; deprojected this is $17.5^\circ$; our estimate of ionization parameter refers to the edge of the cone. The inner and outer filaments have position angle differences with respect  to the milliarcsecond jet of $4^\circ$ and $12^\circ$ respectively. Since we are only determining indicative numbers, we assume that the inclination of the cloud direction to the line of sight is the same as the inclination of the jet to the line of sight (i.e. $\theta_{\rm cl} = \theta_{\rm obs}$) and we calculate the angle between the jet and the cloud direction, $\psi_{\rm cl}$, using equation~(\ref{e:psi_cl}).

The cloud densities assumed for the inner filaments are the same as the estimates of \citet{morganti91a}; the other densities are nominal but the resulting ionization parameters can easily be scaled for other densities.

For the spectra that are observed in Centaurus~A the required ionization parameters are of order a few $\times 10^{-3}$ to $10^{-2}$. Indeed \citet{morganti91a} estimated a reference value from their variable density models for the inner filaments of $5 \times 10^{-3}$. Our estimates for the ionization cone and the inner filaments are in this range for jet Lorentz factors  of four and five but not for the lower Lorentz factor of two. If the outer filaments are to fall in this range the density would need to be an order of magnitude lower. Hence the estimates of jet parameters derived from models of the blue cloud are consistent with photoionization of the \citet{sharp10a} ionization cone. However, the situation is not as clear for the outer filaments.

\acknowledgements We thank Dr. Rob Sharp for informative discussions on Centaurus~A. This work was supported by  the Australian Research Council Discovery Project DP0664434. NN acknowledges the support by the DFG cluster of excellence \emph{Origin and Structure of the Universe}. GVB thanks the European Southern Observatory, where this work was initiated, for its hospitality.


\begin{appendix}

\section{RELATIONSHIP BETWEEN OBSERVED EMISSION AND BEAMED PHOTOIONIZING FLUX}

We often see high energy X-ray and $\gamma$-ray emission from the cores of active galaxies, which is interpreted as emission from a relativistic jet. Potentially, this emission can ionize clouds along the direction of the jet where it is more highly beamed. However, as a result of the beaming pattern of the emission, such clouds see a different intensity to that at the observer. In this Appendix we provide a ready way of estimating the ionizing photon density given parameters such as the Lorentz factor of the jet, the directions of the observer and cloud and the respective distances of cloud and observer. 

\subsection{Ionizing Photon Density}

We use the following symbols and definitions:
\begin{enumerate}
 \item $I(\nu)$ is the specific intensity of a ray propagating from a region (blob) in the jet through the cloud.
 \item $\nu_0 \doteq 3.28 \times 10^{15} \> \rm Hz$ is the frequency of the Lyman limit
 \item $d\Omega$ is the elementary solid angle of rays emanating from the relativistic blob and intersecting the cloud
\item The rest frame of the blob is referred to with a prime.
\item $j^\prime(\nu^\prime)$ is the (isotropic) emissivity in the blob rest frame.
\item $x^\prime$ is the path length along a ray in the blob rest frame.
\item $dA$ is the element of area, in the cloud frame, of the surface $\Sigma$ bounding the relativistically moving plasma; $dA^\prime$
and $\Sigma^\prime$ are the corresponding quantities in the blob rest frame. As spelled out in \cite{lind85a} $\Sigma$ and $\Sigma^\prime$ are congruent; $\Sigma$ is $\Sigma^\prime$ rotated by relativistic aberration and  $dA=dA^\prime$.
\item $d \Omega^\prime$ is the elementary solid angle in the blob rest frame.
\item $\beta = v/c$ for the moving plasma; $\Gamma = (1-\beta^2)^{-1/2}$ is the Lorentz factor. 
\item $\psi_{\rm cl}$ is the angle of the ray passing through the center of the cloud measured with respect to the direction of the jet axis.
\item $\delta_{\rm cl} = \Gamma^{-1} (1-\beta \cos \psi_{\rm cl})^{-1}$ is the Doppler factor of the emitting jet plasma as viewed from the cloud.
\item The frequency, $\nu$, in the cloud frame and the frequency, $\nu^\prime$ in the blob rest frame, are related by $\nu = \delta_{\rm cl} \, \nu^\prime$.
 \item $D_{\rm cl}$ is the distance of the cloud from the relativistic blob. We assume that $D_{\rm cl}$ is large compared to the dimensions of the emitting region.
 \item $dV^\prime$ is the elementary volume of the relativistic blob.
\item $D_A$ is the angular diameter distance of the galaxy containing the relativistic jet.
\item $\delta_{\rm obs} = \Gamma^{-1} (1 - \beta \cos \theta_{\rm obs})^{-1}$ is the Doppler factor of the source along a ray to the observer; $\theta_{\rm obs}$ is the direction of the observer with respect to the blob's velocity.
\item The frequency $\nu^\prime$ of a photon emitted in the jet rest frame and the observed frequency 
$\nu_{\rm obs}$ are related by $\nu_{\rm obs} = \delta_{\rm obs} (1+z)^{-1} \nu^\prime$, where $z$ is the redshift of the galaxy.
\end{enumerate}

The number density of ionizing photons (per unit frequency and total) are, respectively:
\begin{eqnarray}
n_{\rm ph}(\nu) &=& \frac {1}{c} \,  \int \frac{I(\nu)}{h \nu} \> d\Omega \\
n_{\rm ph} &=& \frac {1}{c} \, \int_{\nu_0}^\infty \left[ \int \frac{I(\nu)}{h \nu} \> d\Omega \right] \> d \nu
\label{N_ph}
\end{eqnarray}
We relate these quantities to the parameters of the relativistically moving emission region as follows.

The intensity in the cloud frame and jet rest frame are related by:
\begin{equation}
I(\nu) = \delta_{\rm cl}^3 \, I^\prime(\nu^\prime)
\end{equation}
where 
\begin{equation}
I^\prime(\nu^\prime) = \int_{\rm Ray} j^\prime(\nu^\prime) \> dx^\prime
\end{equation}
and the integral is along a ray through the relativistically moving emitting region $\cal R$.
That is,
\begin{equation}
I(\nu) = \delta_{\rm cl}^3 \, \int_{\rm Ray}  j^\prime(\delta_{\rm cl}^{-1} \nu) \> dx^\prime
\end{equation}
Hence, the number density of photons per unit frequency is
\begin{equation}
n_{\rm ph}(\nu) = 
\frac {1}{c} \,  \delta_{\rm cl}^3 \left[ \int_{\rm Ray} \frac{j^\prime(\delta_{\rm cl}^{-1} \nu)}{h \nu} dx^\prime \right] 
d\Omega
\end{equation}

The number density of ionizing photons in the cloud frame may be determined by integration over the volume of the relativistic emitting region $\cal R$. 
Since $d\Omega = dA/D_{\rm cl}^2 = dA^\prime/D_{\rm cl}^2$, then $dx^\prime d\Omega = dx^\prime dA^\prime/ D_{\rm cl}^2 = dV^\prime/D_{\rm cl}^2$ and
\begin{equation}
n_{\rm ph}(\nu) = \frac {1}{c} \, \frac{\delta_{\rm cl}^3}{D_{\rm cl}^2} \, 
\int_{\cal R} \frac{j^\prime(\delta_{\rm cl}^{-1} \nu)}{h \nu} \>dV^\prime .
\label{Nph_nu}
\end{equation}
The integral over solid angle and path length has been replaced by an integral over comoving volume. We integrate over frequency to obtain the total number of ionizing photons. Thus,
\begin{equation}
n_{\rm ph} = \frac {1}{c} \,  \frac{\delta_{\rm cl}^3}{D_{\rm cl}^2} \, \int_{\nu_0}^\infty
\left[ \int_{\rm Blob} \frac{j^\prime(\delta_{\rm cl}^{-1} \nu)}{h \nu} dV^\prime \right]
\, d \nu
\label{N_tot}
\end{equation}

\subsection{Flux Density at Observer}

For completeness, and in order to relate observed flux density to ionizing photon density we repeat the analysis in \cite{lind85a} in the following expression for the flux density. This enables us to estimate the volume integral of the emissivity which appears in the expressions for the ionizing photon density, equations~(\ref{Nph_nu}) and (\ref{N_tot}). 

The flux density of the blob, at frequency $\nu_{\rm obs}$, as measured by the observer is:
\begin{equation}
F(\nu_{\rm obs}) = \int I_{\rm obs}(\nu_{\rm obs}) \> d \Omega
\end{equation}
As in previous analyses \cite[e.g.][]{lind85a,begelman84a} we determine the observed flux density in two stages, from blob to a point well outside the blob in the galaxy rest frame and from galaxy rest frame to observer. The first stage involves relativistic effects; the second stage cosmological effects. The result for the observed flux density is:
\begin{equation}
F(\nu_{\rm obs}) = \frac {1}{D_A^2} \left( \frac {\delta_{\rm obs}}{1+z} \right)^3 \int_{\cal R} j^\prime 
\left (\frac {(1+z) \nu_{\rm obs}}{\delta_{\rm obs}} \right) \> dV^\prime
\label{F_nu}
\end{equation}
where $j^\prime(\nu^\prime)$ is the emissivity in the jet rest frame.

\subsection{Using the Observed Flux Density to Estimate the Ionizing Photon Density}

We now use the observed flux to determine the volume integral of the emissivity and then the ionizing photon density. The combination 
$\delta_{\rm obs} (1+z)^{-1} \delta_{\rm cl}^{-1} \nu$ which appears in the following reflects the fact that a photon which intersects the cloud with frequency 
$\nu$, originates from the blob rest frame with frequency $\delta_{\rm cl}^{-1} \nu$. Photons emitted with this frequency reach the observer with frequency $\delta_{\rm obs} (1+z)^{-1} \delta_{\rm cl}^{-1} \nu$.

From equation~(\ref{F_nu}) for the flux density, we have:
\begin{equation}
\int_{\cal R} j^\prime(\nu^\prime) \> dV^\prime = D_A^2 \left( \frac {1+z}{\delta_{\rm obs}}\right)^3 \> F\left(\frac{\delta_{\rm obs}}{1+z} \nu^\prime\right)
\label{emissivity}
\end{equation}

When we insert this expression into equation~(\ref{Nph_nu}) for the number density of ionizing photons per unit frequency we obtain:
\begin{equation}
n_{\rm ph}(\nu) =  \frac {1}{c} \, \left( \frac {D_{\rm A}} {D_{\rm cl}}\right)^2 \, \left(\frac {(1+z) \delta_{\rm cl}}{\delta_{\rm obs}} \right)^3 \,
\, \frac {F(\delta_{\rm obs} (1+z)^{-1} \delta_{\rm cl}^{-1} \nu)}{h \nu}
\end{equation}
and the number density of ionizing photons is:
\begin{equation}
n_{\rm ph} =  \frac {1}{c} \, \left( \frac {D_{\rm A}} {D_{\rm cl}}\right)^2 \, 
\left(\frac {(1+z) \delta_{\rm cl}}{\delta_{\rm obs}} \right)^3 \,
\, \int_{\nu_0}^\infty \frac {F(\delta_{\rm obs} (1+z)^{-1} \delta_{\rm cl}^{-1} \nu_0)}{h \nu} \> d\nu
\end{equation}
We now change the integration frequency to the observer's frequency $\nu_{\rm obs} = \delta_{\rm obs} (1+z)^{-1} \delta_{\rm cl}^{-1} \nu$ to obtain:
\begin{equation}
n_{\rm ph} = \frac {1}{c} \, \left( \frac {D_A}{D_{\rm cl}} \right)^2 \, 
\left( \frac {(1+z) \delta_{\rm cl}}{\delta_{\rm obs}} \right)^3  \>
\int_{\delta_{\rm obs} (1+z)^{-1} \delta_{\rm cl}^{-1} \nu_0}^\infty \, 
\frac {F(\nu_{\rm obs})}{h\nu_{\rm obs}} \> d\nu_{\rm obs}
\end{equation}
We now estimate the factor $D_A/D_{\rm cl}$ appearing in this equation. Let the angular separation 
of the cloud from the nucleus in the plane of the sky be $\xi_{\rm cl}$ and the projected linear separation be $D_{\rm cl,p}$. Then, referring to Figure~\ref{f:Geometry} for the source geometry, the polar angle of the ray intersecting the center of the cloud is $\theta_{\rm cl} = \pi - \theta_{\rm obs} - \psi_{\rm cl}$ and the projected cloud distance from the core is
\begin{equation}
D_{\rm cl,p} = D_A \xi_{\rm cl} = D_{\rm cl} \, \sin \theta_{\rm cl}  = 
D_{\rm cl} \, \sin (\theta_{\rm obs} + \psi_{\rm cl})
\label{D_cl}  
\end{equation}
Hence, 
\begin{equation}
\frac {D_A}{D_{\rm cl}} = \frac {\sin (\theta_{\rm obs} + \psi_{\rm cl})}{\xi_{\rm cl}}.
\end{equation}

Thus, the number density of ionizing photons at the cloud is:
\begin{equation}
n_{\rm ph} = \frac {1}{c} \, \frac {\sin^2 (\theta_{\rm obs} + \psi_{\rm cl}) }{\xi_{\rm cl}^2} \,  
\left( \frac {(1+z) \delta_{\rm cl}}{\delta_{\rm obs}} \right)^3  \>
\int_{\delta_{\rm obs}(1+z)^{-1} \delta_{\rm cl}^{-1} \nu_0}^\infty \, \frac {F(\nu_{\rm obs})}{h \nu_{\rm obs}} \> d\nu_{\rm obs}
\end{equation}

This form makes it straightforward to use the observed ionizing spectrum to evaluate the photon density at the cloud as a function of the jet Lorentz factor and the angular direction of the rays intersecting the cloud.

The angular parameter $\psi_{\rm cl}$ depends upon the angle of inclination $\theta_{\rm obs}$. Consider a cloud which is adjacent to the jet and assume that it is approximately spherical. The angle between a ray through the cloud center and the jet axis is determined by the sum of the jet and cloud radii and the distance of the cloud from the core. The jet and cloud radii are independent of the angle of inclination but the distance from the core is affected by projection. Let 
$\Phi_{\rm cl} \approx  0.35^{\prime\prime}$ and $\Phi_{\rm jet} \approx  0.35^{\prime\prime}$ be the observed angular diameter of the cloud and jet respectively. Then the angle, $\psi_{\rm cl}$ between the ray through the center of the cloud and the jet axis is given by
\begin{equation}
\psi_{\rm cl} = \tan^{-1} \left[ \frac {\Phi_{\rm cl} + \Phi_{\rm jet}}{2 \, \xi_{\rm cl}} \, \sin  \theta_{\rm obs} \right]
\approx  \frac {\Phi_{\rm cl} + \Phi_{\rm jet}}{2 \, \xi_{\rm cl}} \, \sin  \theta_{\rm obs}
\label{e:psi_cl}
\end{equation}
where, as above, $\xi_{\rm cl}$ is the projected angular separation of the cloud from the core. In our calculations we round up the value of $\psi_{\rm cl}$  to the nearest degree.

\end{appendix}

\end{document}